\documentclass{article}
\usepackage{bm}
\begin{document}

\title{Neutrino masses from a cobimaximal neutrino mixing matrix} 
\author{Asan Damanik\footnote{E-mal: asandamanik11@gmail.com}\\ {\it Department of Physics Education, Sanata Dharma University}\\ {\it Kampus III USD Paingan Maguwoharjo Sleman Yogyakarta, Indonesia}}
\date{}

\maketitle

\abstract{Recently, we have a confidence that neutrino has a tiny mass and mixing does exist among neutrino flovors as one can see from experimental data that reported by many collaborations.  Based on experimental data that flavor mixing does exist in neutrino sector which imply that all three mixing angles are nonzero, we derive the neutrino mass matrix from a cobimaximal neutrino mixing matrix.  We also evaluate the prediction of neutrino mass matrix with texture zero from a cobimaximal neutrino mixing matrix on  neutrino masses and effective Majorana mass. By using the advantages of experimental data, the obtained neutrino masses are $m_{1}=0.028188$ eV, $m_{2}=0.029488$ eV, and $m_{3}=0.057676$ eV,  and the effective Majorana mass is $\left<m_{\beta\beta}\right>=0.09896$ eV that can be tested in future neutrinoless double beta decay experiments}

\section{Introduction}
One of the main problem in neutrino physics till today is the  explicit form of the neutrino mixing matrix which can accommodate the recent experimental data.  We have already known three well-known neutrino mixing matrices i.e. bimaximal mixing, tribimaximal mixing, and democratic mixing, but now all of the three well-known mixing matrices are inapproriate anymore when confronted to the recent experimental data, especially a nonzero mixing angle $\theta_{13}$ as reported by T2K \cite{T2K} and Daya Bay \cite{Daya} colloaborations.  In order to obtain a mixing matrix which can proceed a consistent predictions with experimental data, Ma \cite{Ma}  proposed a new mixing matrix which is knowm as cobimaximal mixing by assumming mixing angle $\theta_{13}\neq 0$, $\theta_{23}=\pi/4$, and the Dirac phase $\delta=\pm \pi/2$.  In \cite{Ma} also claimed that the cobimaximal neutrino mixing matrix is achieve rigourously in a renormalizable model of radiative charged-lepton and neutrino masses.

In this paper, we evaluate the neutrino mass matrix which is obtained from cobimaximal neutrino mixing matrix.  The obtained neutrino mass matrix is used to predict the neutrino masses and an effective neutrino mass that can be tested in future neutrinoless double beta decaya experiments.  In section II we derive the neutrino mass matrix with neutrino mixing matrix is a cobimaximal mixing.  In section III, we evaluate the power prediction of neutrino mass matrix with texture zero on neutrino masses and effective Majorana mass of neutrinoless double beta decay.  Finally, section IV is devoted for conclusions.

\section{Neutrino mass matrix from a cobimaximal neutrino mixing matrix}
Theoretically, neutrino flavor eigenstates ($\nu_{e},\nu_{\mu},\nu_{\tau}$) are related to neutrino mass eigenstates ($\nu_{1}, \nu_{2}, \nu_{3}$) via a neutrino mixing matrix $V$ as follow
\begin{eqnarray}
\nu_{\alpha}=V_{\alpha\beta}\nu_{\beta},\label{1}
\end{eqnarray}
where the indexes $\alpha=e,\mu,\tau$, $\beta=1,2,3$ , and $V_{\alpha\beta}$ are the elements of the neutrino mixing $V$.  The  standard parameterization of the mixing matrix ($V$) read \cite{Harrison}:
\begin{eqnarray}
V=\bordermatrix{& & &\cr
&c_{12}c_{13} &s_{12}c_{13} &s_{13}e^{-i\delta}\cr
&-s_{12}c_{23}-c_{12}s_{23}s_{13}e^{i\delta} &c_{12}c_{23}-s_{12} s_{23}s_{13}e^{i\delta}&s_{23}c_{13}\cr
&s_{12}s_{23}-c_{12}c_{23}s_{13}e^{i\delta} &-c_{12}s_{23}-s_{12}c_{23}s_{13}e^{i\delta} &c_{23}c_{13}}
 \label{V}
\end{eqnarray}
where $c_{ij}$ is the $\cos\theta_{ij}$, $s_{ij}$ is the $\sin\theta_{ij}$, $\theta_{ij}$ are the mixing angles, and $\delta$ is the Dirac CP-violating phase.

If we use the value of mixing angle $\theta_{23}=\pi/4$,  and the Dirac phase $\delta=\pi/2$, then the neutrino mixing matrix in Eq. (\ref{V}) has the following form
\begin{eqnarray}
V=\bordermatrix{& & &\cr
&c_{12}c_{13} &s_{12}c_{13} & is_{13}\cr
&-\frac{\sqrt{2}}{2}\left(s_{12}-ic_{12}s_{13}\right) &\frac{\sqrt{2}}{2}\left(c_{12}+is_{12}s_{13}\right) &\frac{\sqrt{2}}{2}c_{13}\cr
&\frac{\sqrt{2}}{2}\left(s_{12}+ic_{12}s_{13}\right) &-\frac{\sqrt{2}}{2}\left(c_{12}-is_{12}s_{13}\right) &\frac{\sqrt{2}}{2}c_{13}}
\label{CB}
\end{eqnarray}
which is known as a cobimaximal neutrino mixing matrix.
In the basis where the charged lepton mass matrix is already diagonalized, the neutrino mass matrix defined by the mass term  in the Lagrangian is given by
\begin{eqnarray}
M_{\nu}=VMV^{T},\label{M1}
\end{eqnarray}          
where $M$  is neutrino mass matrix in mass basis
\begin{eqnarray}
M=\bordermatrix{& & &\cr
& m_{1} & 0 & 0\cr
& 0 & m_{2} & 0\cr
& 0 & 0 & m_{3}}
\label{MM}
\end{eqnarray}
By inserting Eqs. (\ref{CB}) and (\ref{MM}) into Eq. (\ref{M1}) we have the following neutrino mass matrix \cite{He}
\begin{eqnarray}
M=\bordermatrix{& & &\cr
&a &b+i\beta &-(b-i\beta)\cr
&b+i\beta &c-i\gamma &d\cr
&-(b-i\beta) &d &c+i\gamma}, \label{M2}
\end{eqnarray}
where
\begin{eqnarray}
a=c_{12}^{2}c_{13}^{2}m_{1}+s_{12}^{2}c_{13}^{2}m_{2}-s_{13}^{2}m_{3},\nonumber\\
b=-\frac{1}{\sqrt{2}}s_{12}^{2}c_{12}^{2}c_{13}^{2}\left(m_{1}-m_{2}\right),\nonumber\\
c=\frac{1}{2}\left((s_{12}^{2}-c_{12}^{2}s_{13}^{2})m_{1}+(c_{12}^{2}-s_{12}^{2}s_{13}^{2})m_{2}+c_{13}^{2}m_{3}\right),\nonumber\\
d=-\frac{1}{2}\left((s_{12}^{2}+c_{12}^{2}s_{13}^{2})m_{1}+(c_{12}^{2}+s_{12}^{2}s_{13}^{2})m_{2}-c_{13}^{2}m_{3}\right),\\
\beta=\frac{1}{\sqrt{2}}s_{13}^{2}c_{13}^{2}\left(c_{12}^{2}m_{1}+s_{12}^{2}m_{2}-m_{3}\right),\nonumber\\
\gamma=-s_{12}c_{12}s_{13}\left(m_{1}-m_{2}\right).\nonumber
\label{abc}
\end{eqnarray}

\section{Neutrino masses and effective Majorana mass}
We are in position to impose the texture zero into neutrino mass matrix of Eq. (\ref{M2}).  By inspecting neutrino mass matrix in Eq. (\ref{M2}) one can see that the realistic neutrino mass matrix with texture zero is by putting the elements of neutrino mass matrix as follow
\begin{eqnarray}
M_{\nu}(1,1)=a=0,\label{Z1}
\end{eqnarray}
and
\begin{eqnarray}
M_{\nu}(2,3)=M_{\nu}(3,2)=d=0,\label{Z2}
\end{eqnarray}
By imposing the texture zero as shown in Eqs. (\ref{Z1}) and (\ref{Z2}), the neutrino mass matrix have the following form
\begin{eqnarray}
M=\bordermatrix{& & &\cr
&0 &b+i\beta &-(b-i\beta)\cr
&b+i\beta &c-i\gamma &0\cr
&-(b-i\beta) &0 &c+i\gamma}. \label{MZe}
\end{eqnarray}

Now, we want to derive relations of neutrino masses as function of mixing angle which then can be used to determine the neutrino masses and its hierarchy by using the experimental data of neutrino oscillation especially the experimental value of squared mass differences.  From Eqs. (\ref{abc}), (\ref{Z1}), and (\ref{Z2}) we can obtain the following relation
\begin{eqnarray}
s_{13}^{2}m_{3}=c_{12}^{2}c_{13}^{2}m_{1}+s_{12}^{2}c_{13}^{2}m_{2},\label{R1}
\end{eqnarray}
and
\begin{eqnarray}
c_{13}^{2}m_{3}=\left(s_{12}^{2}+c_{12}^{2}s_{13}^{2}\right)m_{2}+\left(c_{12}^{2}+s_{12}^{2}s_{13}^{2}\right)m_{2},\label{R2}
\end{eqnarray}
which then proceed
\begin{eqnarray}
m_{3}=m_{1}+m_{2}.\label{MR}
\end{eqnarray}
The Eq. (\ref{MR}) is our new neutrino mass relation when the neutrino mass matrix obtained from a cobimaximal mixing constrained by two texture zero as shown in Eqs. (\ref{Z1}) and (\ref{Z2}).

From Eq. (\ref{MR}) we can have the following relation
\begin{eqnarray}
m_{3}^{2}-m_{2}^{2}=m_{1}^{2}+2m_{2}m_{1},
\end{eqnarray}
or
\begin{eqnarray}
m_{1}^{2}+2m_{2}m_{1}-\Delta m_{32}^{2}=0,\label{SQR}
\end{eqnarray}
where $\Delta m_{32}^{2}=m_{3}^{2}-m_{2}^{2}$ is the squared mass difference of atmospheric neutrino.  The global analysis of squared mass difference for atmospheric neutrino and solar neutrino read \cite{GG}
\begin{eqnarray}
\Delta m_{32}^{2}=2.457\times 10^{-3} \rm{eV^{2}},\label{AN}
\end{eqnarray}
and
\begin{eqnarray}
\Delta m_{21}^{2}=7.50\times 10^{-5} \rm{eV^{2}},\label{SN}
\end{eqnarray}
respectively.

If we insert the value of squared mass difference of Eq. (\ref{AN}) into Eq. (\ref{SQR}), then we have
\begin{eqnarray}
m_{1}=-m_{2}+0.001\sqrt{m_{2}^{2}+2457}.\label{m1m}
\end{eqnarray}
From Eqs. (\ref{SQR}), (\ref{AN}), (\ref{SN}), and (\ref{m1m}) we can have the neutrino masses as follow
\begin{eqnarray}
m_{1}=0.028188 \rm{eV},\nonumber\\
m_{2}=0.029488 \rm{eV},\\
m_{3}=0.057676 \rm{eV}.\label{MM}
\end{eqnarray}
It is apparent from Eq. (\ref{MM}) that hierarchy of neutrino mass is a normal hierarcy.

After we have known the neutrino masses, we then calculate the prediction of cobimaximal mixing on effective Majorana mass because the effective Majorana mass is a parameter of interest in neutrinoless double beta decay.  The effective Majorana mass $\left<m_{\beta\beta}\right>$ is a combination of neutrino mass eigenstates and the neutrino mixing matrix terms as follow \cite{Benato}
\begin{eqnarray}
\left<m_{\beta\beta}\right>=\left|\Sigma V_{ei}^{2}m_{i}\right|,\label{EM}
\end{eqnarray}
where $V_{ei}$ is the  {\it{i}}-th element of the first row of neutrino mixing matrix, and $m_{i}$ is the {\it{i}}-th of the neutrino mass eigenstate.  From Eqs. (\ref{CB}) and (\ref{EM}) we then have an effective Majorana mass
\begin{eqnarray}
\left<m_{\beta\beta}\right>=\left|c_{12}^{2}c_{13}^{2}m_{1}+s_{12}^{2}c_{13}^{2}m_{2}-s_{13}^{2}m_{3}\right|.\label{EMM}
\end{eqnarray}

If we take the central values of mixing angle $\theta_{13}=9^{0}$ \cite{Daya} and $\theta_{12}=35^{0}$ \cite{GG} and the neutrino masses as shown in Eq. (\ref{MM}), then we have the effective Majorana mass as follow
\begin{eqnarray}
\left<m_{\beta\beta}\right>=0.09896 \rm{eV},
\end{eqnarray}
that can be tested in future neutrinoless double beta decay experiments.

\section{Conclusions}
We have use a cobimaximal neutrino mixing matrix to obtain a neutrino mass matrix.  The obtained neutrino mass matrix is constrained by two texture zero i.e. $M_{\nu}(1,1)=0$ and $M_{\nu}(2,3)=M_{\nu}(3,2)=0$ and by using the advantages of the experimental data of squared mass difference, then we can obtain neutrino masses in normal hierarchy i.e. $m_{1}=0.028188 \rm{eV}$, $m_{2}=0.029488 \rm{eV}$, and $m_{3}=0.057676 \rm{eV}$.  By using the central values of reported  experimental mixing angle $\theta_{13}$ and $\theta_{23}$ and obtained neutrino masses, the effective Majorana mass is $\left<m_{\beta\beta}\right>=0.09896 \rm{eV}$ that can be tested in future neutrinoless double beta decay experiments.

\end{document}